\documentclass[twocolumns,aps,prc]{revtex4}

\usepackage{epsfig}
\usepackage{graphicx}

\begin{document}
\title{Open questions and perspectives for hadron electromagnetic form factors in space-like and time-like regions.}

\author{E. Tomasi-Gustafsson}

\email{etomasi@cea.fr}
\affiliation{\it DAPNIA/SPhN, CEA/Saclay, 91191 Gif-sur-Yvette Cedex,
France }

\date{\today}
\pacs{25.30.Bf, 13.40.-f, 13.40.Gp}

\begin{abstract}
An overview of recent data on electromagnetic hadron form factors is presented and discussed in terms of selected nucleon models in space-like and time-like regions. Model independent properties connecting scattering and annihilation channels are presented and give a deeper understanding of the underlying physics. Radiative corrections are discussed. The interest of polarization observables with respect to model predictions and reaction mechanism (one photon versus two photon exchange) is highlighted. 
\end{abstract}
\maketitle

\section{Introduction}

The electromagnetic structure of the nucleon is traditionally described in terms of form factors (FFs), which are related to the current and magnetic distributions. In a $P$ and $T$ invariant theory, a particle of spin $S$ has $2S+1$ form factors: a nucleon, proton or neutron, is described by two FFs, which are different. FFs are measurable quantities, experimentally related to the differential cross section and polarization observables, and calculable by the nucleon models. 

Elastic electron hadron scattering is considered the most direct way to access to FFs, which contain the information on the ground state of the hadron. The interaction is assumed to occur through one photon exchange. In this case, FFs are real functions of one variable, the four momentum squared of the virtual photon, $t=-Q^2<0$. Annihilation processes, as $e^++e^-\leftrightarrow N+\bar N$, also give access to the nucleon FFs, but in another kinematical region, where FFs are complex functions of the momentum transfer squared, $q^2=-Q^2>0$. This is called the time-like (TL) region, as the time component of the four vector $q$ is larger than the space component, in contrast to the scattering region, called the space-like (SL) region. 

The first measurements on the proton in the SL region \cite{Ho62} were rewarded by the Nobel prize to R. Hofstadter in 1964. The formalism of FFs, i.e. the expression of the unpolarized cross section as a function of FFs, was firstly derived in Ref. \cite{Ro50}:
$$
\displaystyle\frac{d\sigma}{d\Omega}=\left \{\epsilon(1+\tau)\left [1+2\displaystyle\frac{E}{m}\sin^2(\theta_e/2)\right ]\displaystyle\frac
{4 E^2\sin^4(\theta_e/2)}{\alpha^2\cos^2(\theta_e/2)}\right \}^{-1}\sigma_{red}(\theta_e,Q^2);
$$
\begin{equation}
\sigma_{red}(\theta_e,Q^2)~=\tau G_M^2(Q^2)+\epsilon G_E^2(Q^2),~\epsilon=[1+2(1+\tau)\tan^2(\theta_e/2)]^{-1},
\label{eq:sigma}
\end{equation}
where $\alpha=1/137$, $\tau=Q^2/(4m^2)$, $Q^2$ is the momentum transfer squared, $m$ is the proton mass, $E$ and $\theta_e$ are the incident electron energy and the scattering angle of the outgoing electron, respectively, and $G_M(Q^2)$ and $G_E(Q^2)$ are the magnetic and the electric proton FFs and are functions of $Q^2$, only.  Measurements of the elastic differential cross section at different angles for a fixed value of $Q^2$ allow $G_E(Q^2)$ and $G_M(Q^2)$ to be determined as the slope and the intercept, respectively, from the linear $\epsilon$ dependence (\ref{eq:sigma}). Many data were collected, improving in precision and/or in the range of momentum transfer squared, the last of unpolarized measurements being recently performed at Jefferson Laboratory \cite{Qa05}. They all suggest the following behavior:
\begin{equation}
G_{E}(Q^2)=G_{M}(Q^2) /\mu_p\simeq G_D,~G_D(Q^2)=(1+Q^2 [\mbox{GeV}^2]/0.71)^{-2}, 
\label{eq:dipole}
\end{equation}
where  $\mu_p=2.79$ is the magnetic moment of the proton and $G_D$ is the dipole function.

Such behavior is, on one side, in agreement with the scaling laws from QCD \cite{Ma73} (which, however, can not predict yet exclusive cross sections), and,  on the other side, with a non relativistic picture of the nucleon, where the charge (magnetic) distribution would have an exponential form (in the Breit system or in a non relativistic approximation, FFs are the Fourier transform of the charge or the magnetic distribution).
As $Q^2$ increases, the precision on the electric FFs gets worse, as the kinematical factor, $\tau$, strongly enhances the magnetic contribution to the unpolarized cross section, and $G_{Mp}$ has been extracted up to a value of the four momentum transfer squared, $Q^2\simeq 31$ GeV$^2$ \cite{Ar75}, while the individual determination of the two FFs, for the proton could be done up to 8.9 GeV$^2$ \cite{And94}.
 
Recently, the polarization method firstly suggested in \cite{Re68}, could be applied \cite{Jo00} due to the availability of high intensity, highly polarized electron beams and polarized targets, and to the optimization of hadron polarimeters in the GeV range. It was shown in \cite{Re68} that the polarization of the elastically scattered proton with a longitudinally polarized electron beam, contains an interference term proportional to the product $G_EG_M$. In the experiment \cite{Jo00} it was proposed to measure the ratio of the longitudinal to transverse proton polarization, in the scattering plane (the normal component vanishes due to parity conservation, if one assumes one photon exchange). These  two components $P_L$ and $P_T$  can be measured simultaneously, and their ratio is directly related to $G_E/G_M$, therefore reducing systematic errors. The data, not only showed a very large precision, but also a large deviation of the ratio $G_E/G_M$ from one, expected from Eq. (\ref{eq:dipole}). This surprising result triggered a large number of theoretical and experimental activity, the two methods being based on the same physics, and no evident problem being found in the experiments.

Although nucleon models predicted such behavior even before the data appeared \cite{Ia73,Ho96}, these data gave rise to a very large debate. Let us enumerate some of the questions:
\begin{itemize}
\item are there models which can reproduce now all FFs for proton and neutron, in SL and TL regions?
\item what are the consequences of these data for the light nuclei, like deuteron or $^3\!He$, which are often described with the help of nucleon FFs? 
\item when the asymptotic region is reached?
\item is the $2\gamma$ exchange the reason of the discrepancy between the FFs measured with the two different methods?
\item which are the experiments or the observables which sign the reaction mechanism?
\end{itemize}
In the next future, measurements are expected in SL region at large momentum transfer for the proton electric FF, and for both the neutron FFs. In the TL region, a program is foreseen at Frascati and Bejing, for $e^+e^-$ annihilation to access proton and neutron FFs, and at FAIR, with $\bar p p$ reactions involving, hopefully, polarization measurements. Special studies are also planned to study the possible presence of the $2\gamma$ exchange mechanism and its interference with the main mechanism, i.e. the one photon exchange, traditionally assumed, at Novosibirsk and JLab. 

The purpose of this work is to focus on some of these issues, looking to a unified description of FFs in TL and SL regions. Model independent statements based on symmetry properties of the strong interaction will ba also recalled.

\section{Overview of the existing results}
 
In Ref. \cite{ETG05} a global analysis of the existing data was done with phenomenological models proposed in the literature for the SL region analytically continued to the TL region. After determining the parameters through a fit on the available data, predictions for the observables which will be experimentally accessible with large statistics, polarized annihilation reactions were given.

Not all the existing models of nucleon FFs can be extrapolated to TL region. We consider predictions of pQCD, in a form generally used as simple fit to experimental data, a model based on vector meson dominance (VMD) \cite{Ia73}, and a third model based on an extension of VMD, with additional terms in order to satisfy the asymptotic predictions of QCD \cite{Lomon}, in the form called GKex(02L).

\begin{figure}[pht]
\begin{center}
\includegraphics[width=16cm]{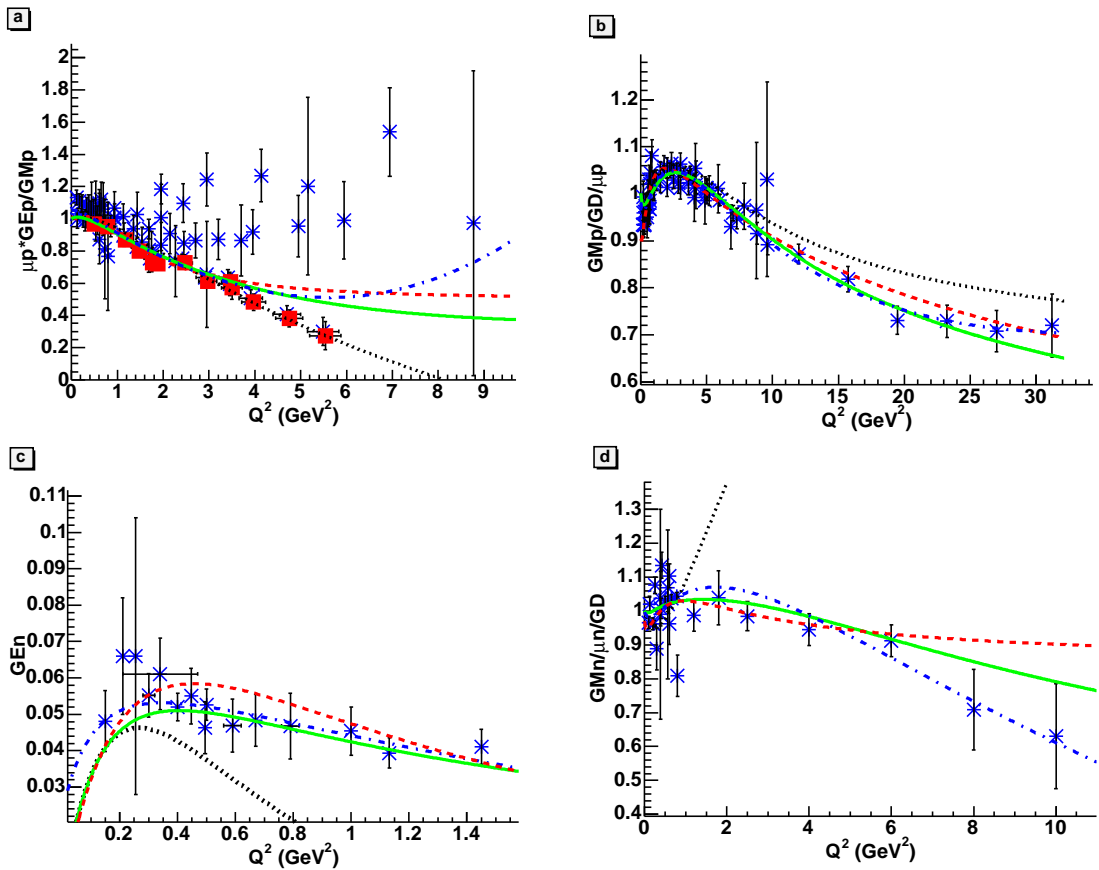}
\caption{\label{fig:fig1} Nucleon Form Factors in SL region: (a) proton electric FF, scaled by $\mu_p G_{Mp}$ (b) proton magnetic FF scaled by $\mu_p G_D$ , (c) neutron electric FF, (d) neutron magnetic FF, scaled by $\mu_n G_D$. The predictions of the models are drawn:  from Ref. \cite{Ia73} (dotted line),  from Ref. \cite{Lomon} (solid line), model from Ref. Ref. \cite{Ho76} (dash-dotted line), from \cite{Bo95} (dashed line). }
\end{center}
\end{figure}

In Fig. \ref{fig:fig1} the world data on the form factors for proton and neutron (top and bottom, respectively) electric (left) and magnetic (right) are reported.  For the electric proton FF, the discrepancy among the data measured with the Rosenbluth methods (stars) and the polarization method (solid squares) appears clearly in Fig. \ref{fig:fig1}a. This problem has widely been discussed in the literature and rises fundamental issues. If the trend indicated by polarization measurements is confirmed at higher $Q^2$ \cite{04108}, not only the electric and magnetic charge distribution in the nucleus are different and deviate, classically, from an exponential charge distribution, but also the electric FF has a zero and becomes eventually negative. This scenario will change our view on the nucleon structure and will favor VMD inspired models like \cite{Du03,Bij04}, which can reproduce such behavior. 

\begin{figure}[pht]
\begin{center}
\includegraphics[width=16cm]{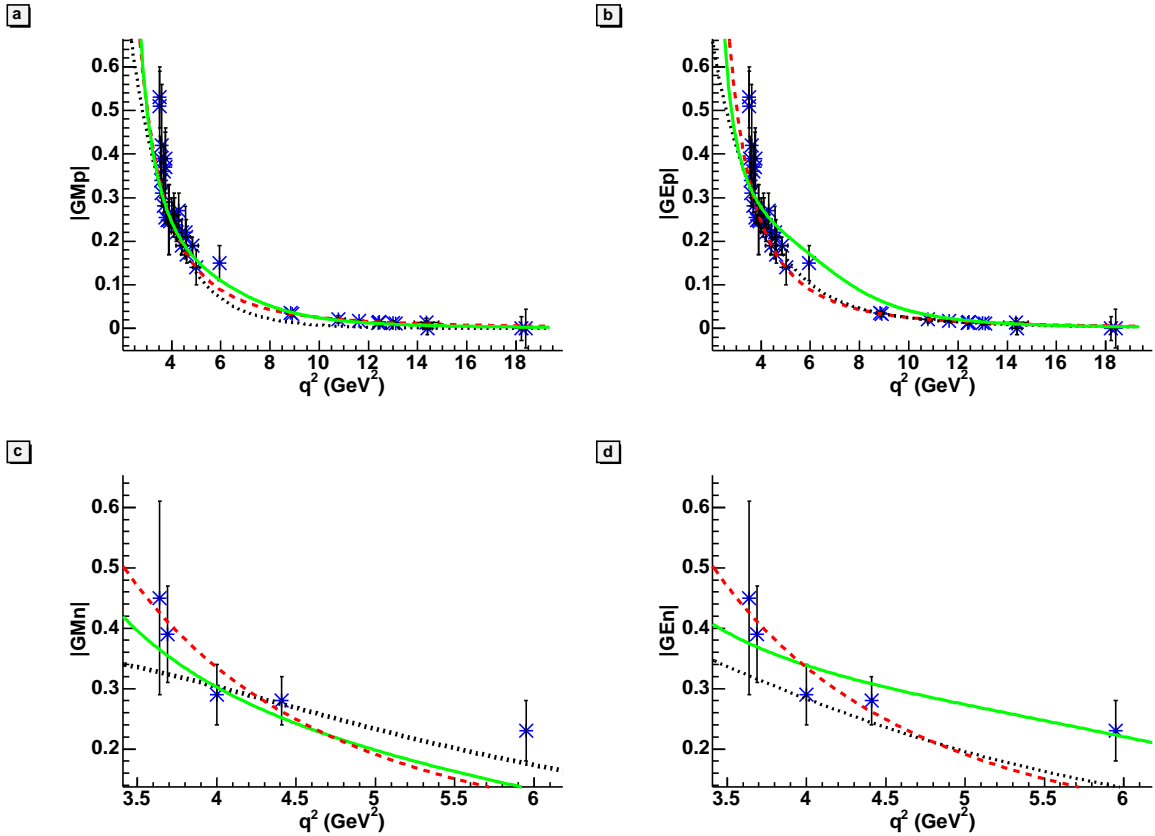}
\caption{\label{fig:fig2} Form Factors in TL region and predictions of the models: pQCD-inspired  (dashed line), from Ref. \cite{Ia73} (dotted line), from Ref. \cite{Lomon} (solid line).}
\end{center}
\end{figure}

The data in the TL region are drawn in Fig. \ref{fig:fig2}a, b  for the proton and in Fig. \ref{fig:fig2}c, d for the neutron. As no separation has been done for electric and magnetic FFs, the data are extracted under the hypothesis that $|G_{EN}|=|G_{MN}|$. Concerning the neutron, the first and still unique measurement was done at Frascati, by the collaboration FENICE \cite{An94}. The models are fitted to the data, assuming that they correspond to $|G_M|$, and the curves in the $|G_E|$ plots should be considered predictions.

Figs. \ref{fig:fig1} and \ref{fig:fig2} show that it is possible to find a satisfactory general representation of all nucleon FFs. The parametrization from Ref. \cite{Ia73} (dotted line) is based on a view of the nucleon as composed by an inner core with a small radius (described by a dipole term) surrounded by a meson cloud. In framework of this model a good global fit in SL region has been obtained with a modification including a phase in the common dipole term. However, the TL region is less well reproduced \cite{Bij04}. The curves drawn in all the figures correspond to the original parameters. The result from  \cite{Lomon} (solid line) gives a good overall parametrization, with parameters not far from those found in the original paper for the SL region only. Phenomenological parametrization, commonly used in the SL region are also shown (from Ref. \cite{Ho76} (dash-dotted line) and from \cite{Bo95} (dashed line)), but their extrapolation to the TL region gives rise to large discontinuities. 

A possible explanation of the fact that FFs are systematically larger in TL region than in SL region (which is true also in the proton case) is the presence of a narrow resonance in the $N\overline{N}$ system, just below the $N\overline{N}$ threshold \cite{Ga96}.  

Several pQCD inspired parametrizations exist for the form factor ratio $F_2/F_1$, which include logarithmic corrections, and have been recently discussed in Ref. \cite{Br03}. However, some of these analytical forms have problems related to the asymptotic behavior \cite{ETG05a}. The Phr\`agmen-Lindel\"of 
theorem \cite{Ti39} gives a rigorous prescription for the asymptotic behavior of analytical functions: 
$\lim_{q^2\to -\infty} F^{(SL)}(q^2) =\lim_{q^2\to \infty} 
F^{(TL)}(q^2)$.
This means that, asymptotically, FFs, have the following constraints: 
1) the TL phase vanishes  and 2) the real part of FFs, 
${\cal R}e  F^{(TL)}(q^2)$, coincides with the 
corresponding value, $F^{(SL)}(q^2)$. These asymptotic properties based on analiticity, are different from the asymptotics properties of FFs, predicted in QCD, which derive from scaling rules and helicity conservation. Therefore, the study of FFs at large $Q^2$ represents a unique tool to understanding these properties of the nucleon electrodynamics.
\section{Polarization observables}

The importance of polarization observables has been quoted above, in connection with the proton electromagnetic FFs. The polarization method \cite{Re68} has been also successfully applied for a precise determination of the neutron FFs, and shows that $G_{En}$ is definitely different from zero (\cite{Se04} and refs therein). 

No polarization experiment has been done in TL region. The interest of such measurements is evident from Fig. 3, where predictions for different observables are shown, for the three models detailed above, which all reproduce satisfactorily the existing unpolarized data. All observables manifest a different behavior, according to the different models. Even the sign can be different for VMD inspired models and pQCD. 

\begin{figure}[pht]
\begin{center}
\includegraphics[width=16cm]{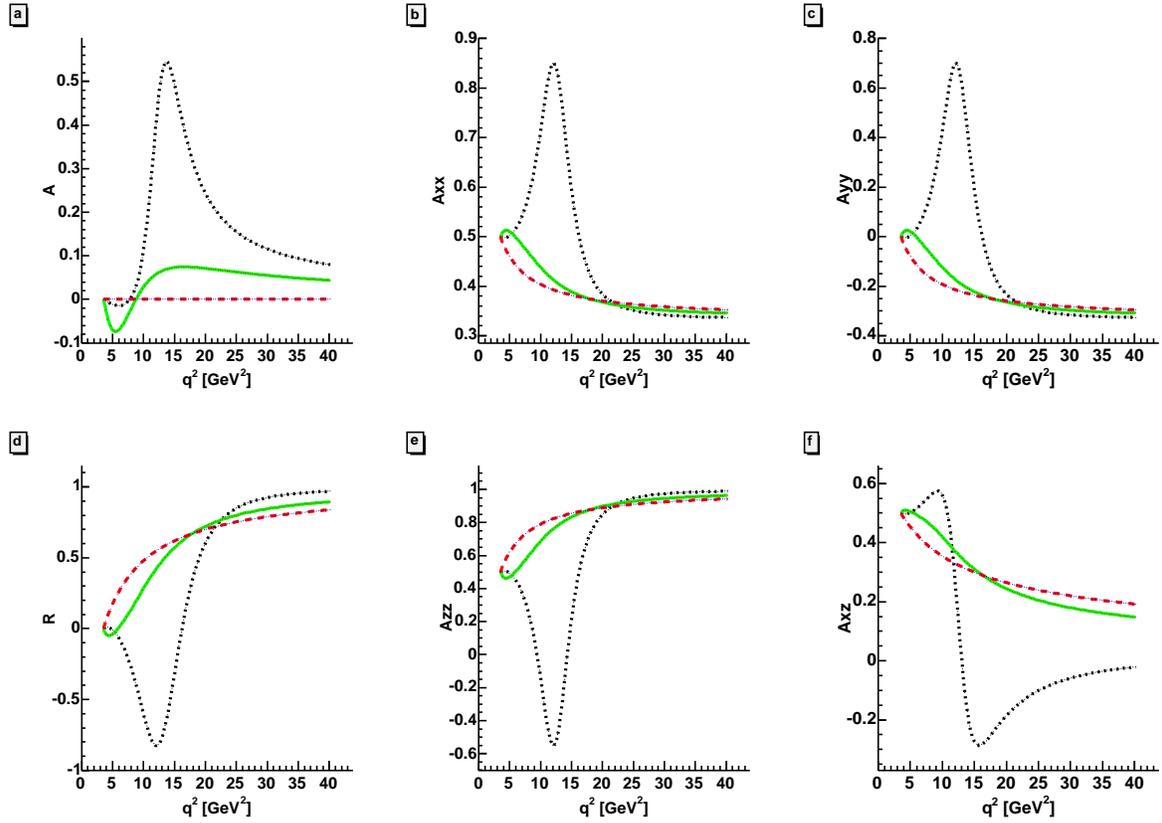}
\caption{\label{fig:fig3} Angular asymmetry and polarization observables, for a fixed value of $\theta=45^0$. Notations as in Fig. \protect\ref{fig:fig2}.}
\end{center}
\end{figure}

The single spin asymmetry for the process $p+\overline{p}\to e^++e^-$, the angular asymmetry  and the double spin observables $A_{ab}$ are  shown in Fig.  \ref{fig:fig3}. Here $a$ and $b=x,y,z$ refer to the $a(b)$ component of the projectile (target) polarization in the CMS system, where the $z$ axis is taken along the direction of the incoming antiproton, the $y$ axis normal to the scattering plane, and the $x$ axis to form a left-handed coordinate system. As shown in Fig. \ref{fig:fig3}, all these observables are, generally, quite large. The model \cite{Ia73} predicts the largest (absolute) value at  $q^2\simeq$ 15 GeV $^2$ for all observables, except $A_{xz}$, which has two pronounced extrema. The fact that single spin observables in annihilation reactions are discriminative towards models, especially at threshold, was already pointed out in Ref. \cite{Dub96}, for the process $e^++e^-\to p+\overline{p}$ on the basis of two versions of a unitary VDM model, and, more recently in \cite{Br03}. The present results, (Fig. \ref{fig:fig3}), for the inverse reaction $p+\overline{p}\to e^++e^-$ confirm this trend and show that experimental data will be extremely useful, particularly in the kinematical region around $q^2\simeq$ 15 GeV $^2$. 

In TL region, polarization data allow to access the phase of FFs. In order to determine the relative phase of FFs, in TL region, the interesting observables are $A_y$, and $A_{xz}$ which contain, respectively, the imaginary and the real part of the  product $G_EG_M^*$.

In Fig. \ref{fig:fig3}d the angular asymmetry is also shown. The electric and the magnetic FFs are weighted by different angular terms, in the expression of the differential cross section. One can define an angular asymmetry, ${\cal R}$, with respect to the differential cross section measured at $\theta=\pi/2$, $\sigma_0$ \cite{ETG01}:
\begin{equation}
\left (\displaystyle\frac{d\sigma}{d\Omega}\right )_0=
\sigma_0\left [ 1+{\cal R} \cos^2\theta \right ],
\label{eq:asym}
\end{equation}
where ${\cal R}$ can be expressed as a function of FFs:
\begin{equation}
{\cal R}=\displaystyle\frac{\tau|G_M|^2-|G_E|^2}{\tau|G_M|^2+|G_E|^2}.
\end{equation}
This observable is also very sensitive to the different underlying 
assumptions on FFs, therefore, a precise measurement of this quantity, which does not require polarized particles, would be very interesting.

\section{Radiative corrections}

\begin{figure}
\begin{center}
\includegraphics[width=12cm]{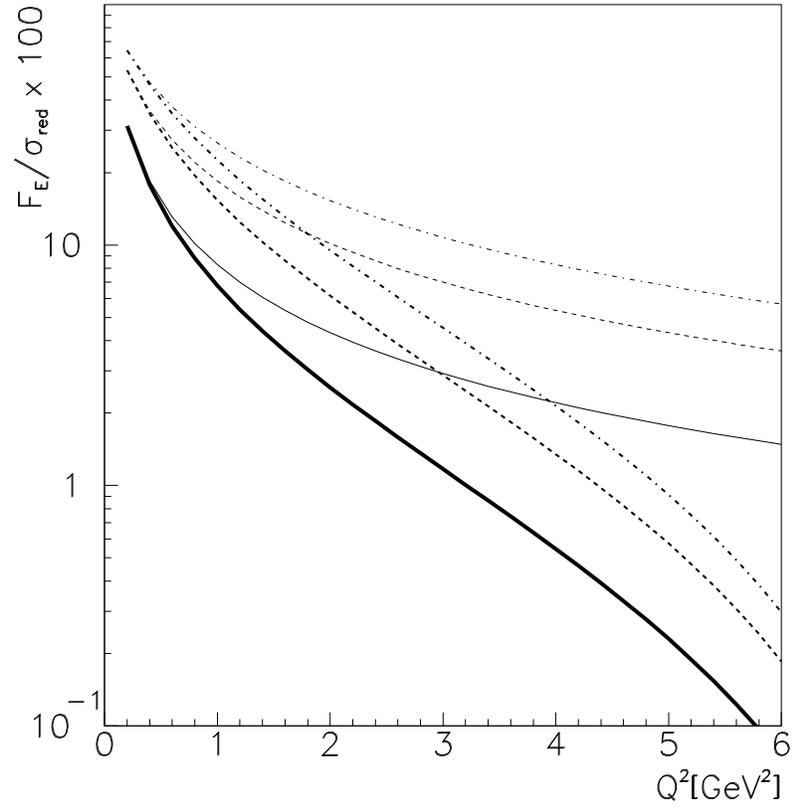}
\caption{\label{Fig:fig2} Contribution of the $G_E(Q^2)$ dependent term to the reduced cross section (in percent) for $\epsilon=0.2$ (solid line), $\epsilon=0.5$ (dashed line), $\epsilon=0.8$ (dash-dotted line), in the hypothesis of FF scaling (thin lines) or following the expectation from polarization data (thick lines).}
\end{center}
\end{figure}

\begin{figure}
\begin{center}
\includegraphics[width=17cm]{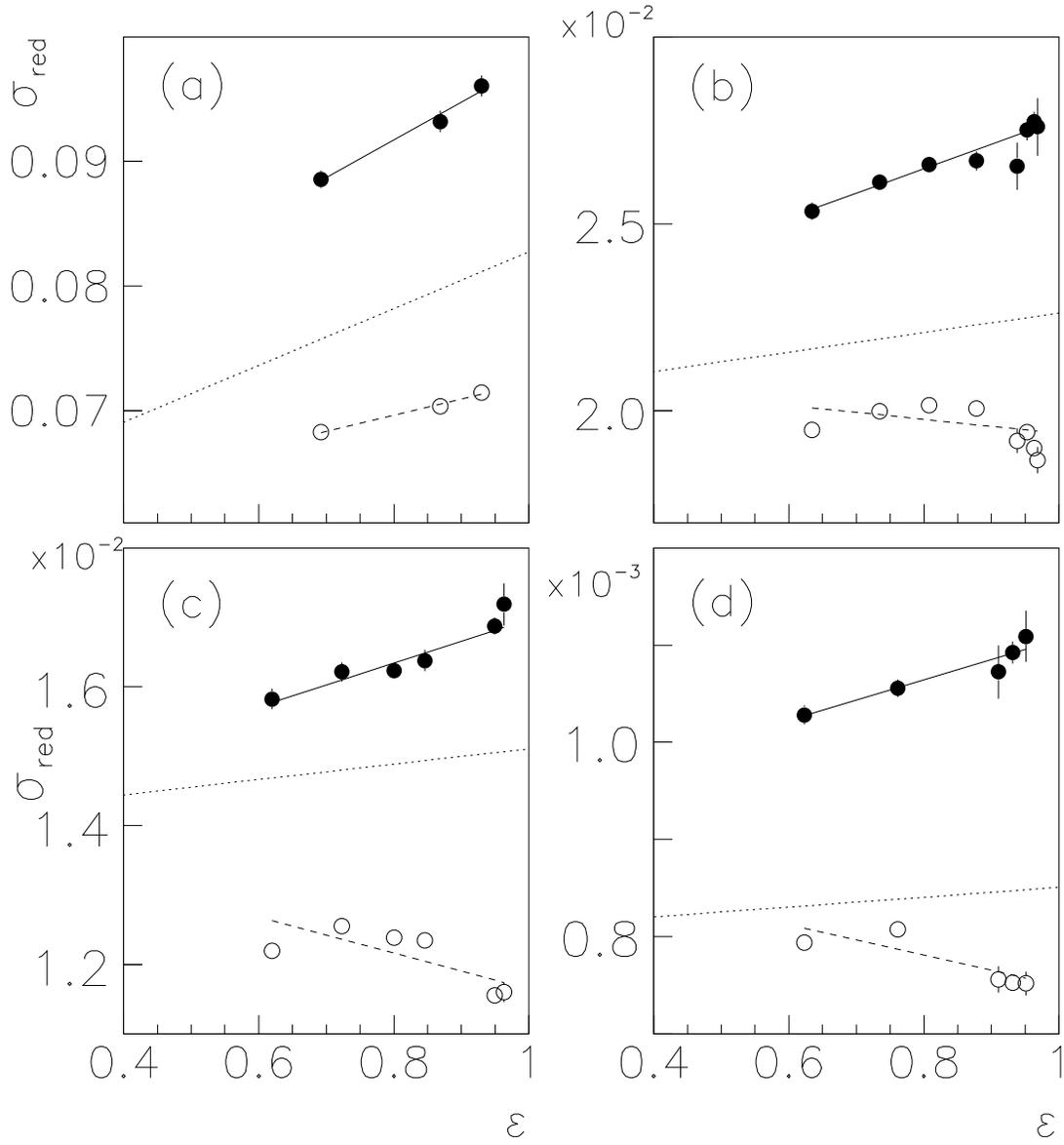}
\caption{\label{fig:sl1} Reduced cross section with (solid circle) and without (open circles) radiative corrections, for $Q^2$=1, 2, 2.5,  and 3 GeV$^2$. Data are from Ref. \protect\cite{Wa94} and two-parameters linear fits (solid and dashed lines). The dotted lines corresponds to the slopes suggested by the polarization data. }
\end{center}
\end{figure}

At large $Q^2$ the Rosenbluth method is not very effective for the extraction of $G_E$, essentially for two reasons: the contribution of the electric term to the cross section becomes very small, as the magnetic part is amplified by the kinematical factor $\tau $, and radiative corrections, which should be applied to the data, can reach 30-40\%. In Fig. \ref{Fig:fig2}, the ratio of the electric part, $F_E=\epsilon G_E^2(Q^2)$, to the reduced cross section is shown as a function of $Q^2$. The different curves correspond to different values of $\epsilon$, assuming FFs scaling (thin lines) or in the hypothesis of the dependence suggested by the polarization method (thick lines). In the second case, one can see that, for example, for $\epsilon=0.2$ the electric contribution becomes lower than $3$\% starting from 2 GeV$^2$. This number should be compared with the absolute uncertainty of the cross section measurement. When this contribution is larger or is of the same order, the sensitivity of the measurement to the electric term is lost and the extraction of $G_E(Q^2)$ becomes meaningless.
The measured elastic cross section is usually corrected by a global factor $C_R$, according to the prescription \cite{Mo69}:
\begin{equation}
\sigma_{red}= C_R\sigma_{red}^{meas}.
\label{eq:sred}
\end{equation}
The factor $C_R$ {\it contains a large $\epsilon$ dependence} and a smooth $Q^2$ dependence,  and it is common to the electric and magnetic parts. At the largest $Q^2$ considered here this factor  getting larger when the resolution is higher. Note that radiative corrections are not applied to the polarization results, as they are assumed to cancel in the polarization ratio \cite{Af02}, although  a complete calculation of all contributions and their interference does not exist. If one makes a linear approximation for the uncorrected data, one  even finds  that the slope of the {\it measured} reduced cross section as a function of $\epsilon$, (which is related to the electric FF squared), can vanish and may even become negative \cite{ETG}. This is shown in Fig. \ref{fig:sl1} for the data of Ref. \cite{Wa94}.  A similar study has been done from all available sets of data and leads to similar results.

The extraction of $G_E$ from the unpolarized data requires a large precision in the calculation of radiative corrections, and in the procedure used to apply to the data. In particular at large $\epsilon$, an overestimate of the radiative corrections, can be source of a large change in the slope. Moreover, the procedure applied to correct the data, induces large systematic errors in the parameters of the Rosenbluth fit, as correlations approach 100\% at large $Q^2$ \cite{Etg05b}.

\begin{figure}
\begin{center}
\includegraphics[width=14cm]{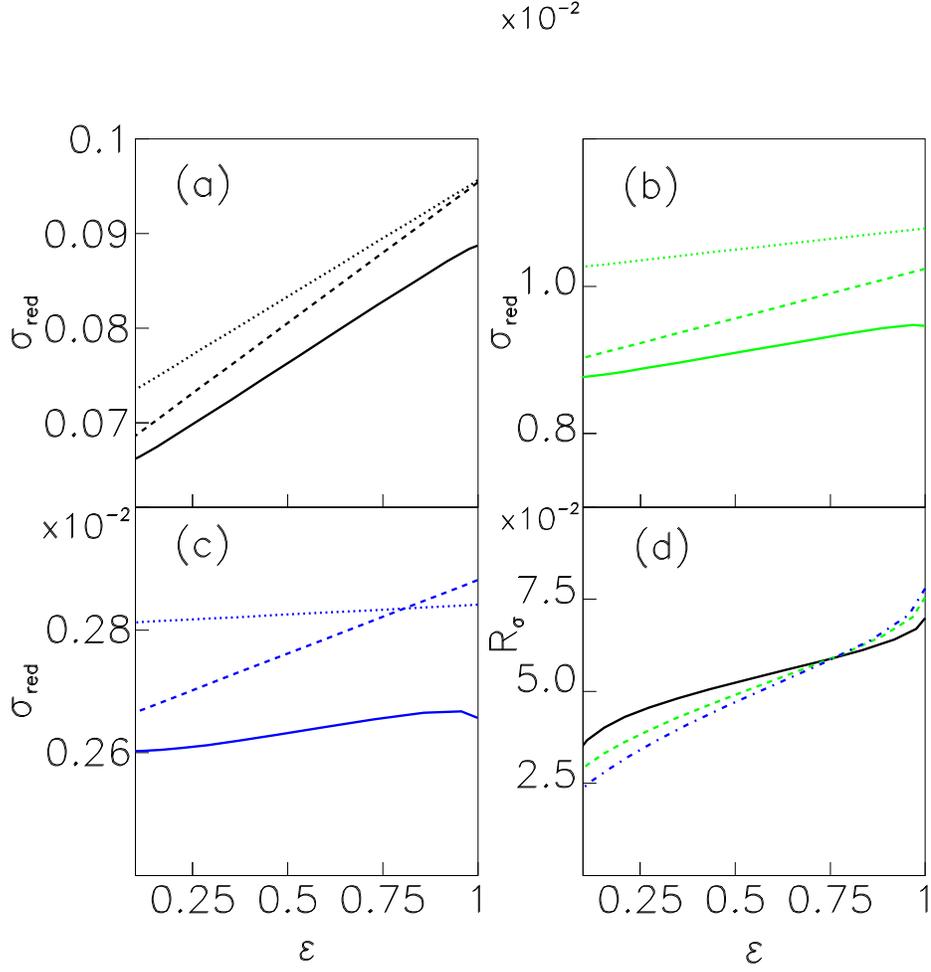}
\caption{\label{fig:fig5}  Reduced cross section for $ep$ elastic scattering as a function of $\epsilon$, at $Q^2$=1 GeV$^2$ (a), 3 GeV$^2$ (b), and  5 GeV$^2$ (c) from the SF method (solid line) and from the Born approximation (dashed line) for dipole parametrization of FFs. The relative difference, $R_{\sigma}$, between these two calculations is shown in (d) for $Q^2$=1 GeV$^2$ (solid line), 3 GeV$^2$ (dotted line), and  5 GeV$^2$ (dash-dotted line). For comparison, the expectation from the polarization data is shown as a dotted line, in (a), (b) and (c). }
\end{center}
\end{figure}

A different calculation of radiative corrections, based on the structure functions method \cite{Ku85}, has been applied to the unpolarized cross section and compared to the unpolarized cross section, calculated in the Born approximation, using in both cases dipole FFs. The results are shown in Fig. \ref{fig:fig5}a, b, c for different $Q^2$ values, $Q^2$=1, 3, and 5 GeV$^2$ (solid lines). For comparison, the Born  reduced cross section assuming dipole FFs is also shown (dashed line), and according to polarization measurements (dotted line). The relative difference between the two calculations is shown in Fig. \ref{fig:fig5}d. It is $\epsilon$ dependent and  reaches 7\% at the largest $\epsilon$ values. The method from \cite{Ku85} gives a different, uncorrelated contribution to the electric and magnetic terms and consequently a different slope, which is smaller and compatible with the result from polarization measurements. Evidently, this result depends quantitatively  on the inelasticity cut in the scattered electron energy spectrum, which has been fixed at 3\% of the elastic peak, according to the resolution of the present experiments \cite{Ku06}.
\section{Two-photon exchange}
All the considerations and results presented above are based on the assumption that the main reaction mechanism is one-photon exchange, the two-photon contribution being suppressed by the order of $\alpha$. However, long ago,  
\cite{Gu73} it was shown that, due to the steep decreasing of FFs with $Q^2$, a mechanism where the momentum transfer is equally shared between the two photons could compensate the simple rule of $\alpha$ counting. However the exact calculation of the $2\gamma$-contribution to the amplitude of the $e^{\pm} p\to e^{\pm} p$-process requires the knowledge of the matrix element for the double virtual Compton scattering, $\gamma^*+N\to\gamma^*+N$, in a large kinematical region of colliding energy and virtuality of both photons, and can not be done in a model independent form.

The reactions $e^{\pm}+ p\to e^{\pm} +p$ and $e^+ +e^-\leftrightarrow p+\bar p$ are connected by crossing symmetry and T-reversal, therefore general properties of the hadron electromagnetic interaction, as  C-invariance and crossing symmetry can be applied and give rigorous and model independent prescriptions for different observables in particular for the differential cross section and for the proton polarization, induced by polarized electrons \cite{Re04}. These concrete prescriptions help in identifying a possible manifestation of the two-photon exchange mechanism. 

Let us consider here the case of $e^{\pm}+ p\to e^{\pm}+ p$ scattering. First of all, assuming electron helicity conservation, the spin structure of the matrix element contains three complex amplitudes, which are functions of two kinematical variables, the total energy $s$ and the momentum transfer $Q^2$, instead of two real functions of $Q^2$, as in case of one photon exchange. Moreover, the amplitudes for electron and positron scattering will be different and their connection with the nucleon FFs not trivial. But these amplitudes
have specific  symmetry properties with respect to the change $x\to -x$ $\left (x=\sqrt{\displaystyle\frac{1+\epsilon}{1-\epsilon}}\right )$. 

Crossing symmetry requires that the two channels $e^-+N\to e^-+N$, in $s$--channel, and $e^++e^-\to N+\overline{N}$, in $t$--channel are described by the same amplitudes. The transformation from $s$- to $t$-channel can be realized by the following substitution: $k_2\to -k_2,~p_1\to -p_1$, 
and for the invariant variables:
$s=(k_1+p_1)^2\to (k_1-p_1)^2,~Q^2=-(k_1-k_2)^2\to -(k_1+k_2)^2=-t.$

Let us consider firstly the one-photon mechanism for $e^++e^-\to p+\overline{p}$. The conservation of the total angular momentum ${\cal J}$  allows one value, ${\cal J}=1$ , and the quantum numbers of the photon: ${\cal J}^P=1^-$, $C=-1$. The selection rules with respect to the C and P-invariances allow two states for
$e^+e^-$ (and $p\overline{p}$):
\begin{equation}
S=1,~\ell=0 \mbox{~and~} S=1,~\ell=2\mbox{~with~} {\cal J}^P=1^-,
\label{eq:tran}
\end{equation}
where $S$ is the total spin and $\ell$ is the orbital angular momentum. As a result the $\theta$-dependence of the cross section for $e^++e^-\to p+\overline{p}$, in the one-photon exchange mechanism is:
\begin{equation}
\displaystyle\frac{d\sigma}{d \Omega}(e^++e^-\to p+\overline{p})\simeq a(t)+b(t)\cos^2\theta,
\label{eq:sig}
\end{equation}
where $a(t)$ and $b(t)$ are definite quadratic contributions of $G_{Ep}(t)$ and
$G_{Mp}(t)$, $a(t),~b(t)\ge 0$ at $t\ge 4m^2$.

Using the kinematical relations:
\begin{equation}
\cos^2\theta=\displaystyle\frac{1+\epsilon }{1-\epsilon}\to 
\displaystyle\frac{\cot^2{\theta_e/2}}{1+\tau}+1
\label{eq:cot}
\end{equation}
between the proton emission scattering angle $\theta$, (in the CMS of $e^++e^-\to p+\overline{p}$) and the angle of the scattered electron $\theta_e$, (in the LAB system for $e^-+p\to e^-+p$), it appears clearly that the one-photon mechanism generates a linear $\epsilon$-dependence (or  $\cot^2{\theta_e/2}$) of the Rosenbluth differential cross section for elastic $eN$-scattering in Lab system.

Let us consider now the $\cos\theta$-dependence of the $1\gamma\bigotimes 2\gamma$-interference contribution to the differential cross section of  $e^++e^-\to p+\overline{p}$. The spin and parity of the $2\gamma$-states
is not fixed, in general, but only a positive value of C-parity, $C(2\gamma)=+1$, is allowed.
An infinite number of  states with different quantum numbers can contribute, and their relative role is determined by the dynamics of the process $\gamma^*+\gamma^*\to  p+\overline{p}$, with both virtual photons.

But the $\cos\theta$-dependence of the contribution to the differential cross section for the $1\gamma\bigotimes 2\gamma$-interference can be predicted on the basis of its C-odd nature:
\begin{equation}
\displaystyle\frac{d\sigma^{(int)}}{d \Omega}(e^++e^-\to p+\overline{p})=\cos\theta[c_0(t)+c_1(t)\cos^2\theta+c_2(t)\cos^4\theta+...],
\label{eq:sig3}
\end{equation}
where $c_i(t)$, $i=0,1..$ are real coefficients, which are functions of $t$,  only. This odd $\cos\theta$-dependence is essentially different from the even $\cos\theta$-dependence of the cross section for the one-photon approximation. One can conclude that the linearity of the Rosenbluth fit is destroyed by the presence of the two photon mechanism.

In spite of the complexity induced by the presence of two photon exchange in the spin structure of the matrix element, it has been shown in Ref. \cite{Re04} that it is still possible to extract FFs, in $ep$ elastic scattering, through the measurement of 5 T-odd or 3 T-even polarization observables, including triple-spin observables. Due to the difficulties of such measerements, and to the implications in all data issued from electroproduction experiments, the experimental observation of this mechanism is necessary. The experiment, planned at Novosibirk, is a precise measurement of the difference of electron positron elastic scattering on the proton, which is due only to the $2\gamma$ contribution, while the sum contains only $1\gamma$ terms (neglecting $2\gamma$ terms squared \cite{Ni05}. 

Other observables sign the presence of $2\gamma$ exchange:

- In $ep$ elastic scattering, the normal component of the proton polarization should vanish in the $1\gamma$ approximation. 

- In the annihilation channel it has been shown that the single spin asymmetry in $\vec p+\overline{p}\to e^++e^-$ \cite{Ga05a}, or the polarization of the proton in $e^++e^-\to p+\overline{p}$ \cite{Ga05b}, would give also a signature of the $2\gamma$ presence, in particular kinematical conditions. For $1\gamma$ exchange, these observables do not vanish in the annihilation region, due to the complexity of FFs, except at $90^\circ$ or in the threshold region, where $G_E=G_M$.

\section{Conclusions and Perspectives}

The field of hadron EM FFs is presently the object of huge experimental and
theoretical investigations, due essentially to the large achievements on the experimental side, that reached a high degree of precision, at large momentum transfer, with the help of polarization techniques.

The main new results based on the polarization method, concern the ratio of the charge and magnetic FFs of the proton, discussed here, and the electric neutron form factor which is small, but does not vanish.

In future, the possibility to extend these measurements at high $Q^2$ and/or  with larger precision, in TL as well as in SL region, has given rise to a number of proposals at different accelerators in the world. The extension of the proton FFs ratio measurement up to $Q^2=$9 GeV$^2$ will show, if the present trend is extrapolated, the presence of a zero in the electric FF. The precise study of the $e^\pm p$ cross section at Novosibirsk will sign the presence of the $2\gamma$ contribution, if any. But even larger progress is expected in the TL region, where data are scarce and no individual determination of the two nucleon FFs has been done yet. In particular new data on the neutron are expected in Frascati, as well as polarization experiments which will allow to determine firstly the relative phase between the complex electric and magnetic FFs. More precise measurements in the threshold region are expected also from Bejing. New techniques using radiative return seem also promising \cite{Czyz}. The region of large $Q^2$ should be investigated, in particular, at FAIR, with the future beams of (polarized) protons and antiprotons \cite{GSI}.
\section{Aknowledgments}
Most part of the work presented here was done in collaboration or inspired by enlightning discussion with M. P. Rekalo. Thanks are due to G. I. Gakh, for a careful reading of the manuscript and useful remarks, and to E.A. Kuraev for suggesting the structure function method, in the calculation of radiative corrections.

{}
\end{document}